\documentclass{article}

\usepackage{arxiv}

\usepackage{xcolor}

\usepackage{amsthm}

\newtheoremstyle{named}{}{}{\itshape}{}{\bfseries}{.}{.5em}{#1 \thmnote{#3}}
\theoremstyle{named}

\usepackage[utf8]{inputenc} 
\usepackage[T1]{fontenc}    
\usepackage{hyperref}       
\usepackage{url}            
\usepackage{booktabs}       
\usepackage{amsfonts}       
\usepackage{nicefrac}       
\usepackage{microtype}      
\usepackage{lipsum,cite}
\usepackage{graphicx,xcolor,amsmath,amssymb}

\graphicspath{{./Figures/}}
\usepackage{float}

\title{Using machine learning methods to predict cognitive age from psychophysiological tests}

\iftrue

\author{
 Daria D. Tyurina\\
  Lobachevsky University, Russia\\
  \texttt{tyurina0410@gmail.com} 
 \And
 Sergey V. Stasenko\\
  Lobachevsky University, Russia\\
  \texttt{stasenko@neuro.nnov.ru} 
   \And
 Konstantin V. Lushnikov\\
 Lobachevsky University, Russia\\
  \texttt{kolushn1@yandex.ru} 
   \And
 Maria V. Vedunova\\
 Lobachevsky University, Russia\\
  \texttt{mvedunova@ya.ru} 
}

\fi

\begin{document}
\maketitle

\begin{abstract}
This study introduces a novel method for predicting cognitive age using psychophysiological tests. To determine cognitive age, subjects were asked to complete a series of psychological tests measuring various cognitive functions, including reaction time and cognitive conflict, short-term memory, verbal functions, and color and spatial perception. Based on the tests completed, the average completion time, proportion of correct answers, average absolute delta of the color campimetry test, number of guessed words in the Münsterberg matrix, and other parameters were calculated for each subject. The obtained characteristics of the subjects were preprocessed and used to train a machine learning algorithm implementing a regression task for predicting a person's cognitive age. These findings contribute to the field of remote screening using mobile devices for human health for diagnosing and monitoring cognitive aging.
\end{abstract}


\section{Introduction}
Cognitive decline associated with aging is an increasingly serious public health problem. By 2050, at least 152.8 million people worldwide are projected to have dementia \cite{nichols2022estimation}. Aging is a complex, natural process that results in varying degrees of cognitive decline \cite{babulal2020advancing}. Some older adults maintain acceptable levels of cognitive function, while others experience varying degrees of cognitive decline that can lead to pathological conditions \cite{cohen2019neuropsychology}. Such decline can significantly impair a person's ability to live independently \cite{grady2012trends}. Therefore, researchers are constantly searching for methods to mitigate or even reverse both the physiological and pathological aspects of cognitive decline \cite{shipstead2012working,soveri2017pattern}. Although treatments for cognitive impairment and dementia exist, significant progress is currently impossible to achieve \cite{cummings2021alzheimer}, so additional methods for both early diagnosis of cognitive ability and interventions are needed to help reduce the personal, social, and economic costs associated with the increasing number of dementia diagnoses \cite{xu2017global}.
However, the problem is that cognitive function is not clearly linked to biological age and physical condition. Some researchers suggest that cognitive ability peaks around age 30 and then gradually declines, while others point to a sharp decline in cognitive function between the ages of 50 and 60 \cite{schroeder2004age, brockmole2013age, logie2009internet,johnson2010working, hartshorne2015does,dobbs1989adult}. Others, conversely, argue that the decline occurs gradually, without a precise age limit \cite{nowak2021altered, thornton2006aging,missonnier2011age,archer2018working}.
Considering that different types of information are processed by different cognitive systems and are susceptible to aging to different degrees \cite{baddeley2012working,salthouse1995aging,logie2011functional,jenkins2000converging,chen2003effects}, we used different tests.

However, when developing a mathematical model for determining cognitive age, it is necessary to consider the "biological reality," which is as follows. The results of psychological and psychophysiological tests exhibit significant variability, which complicates their use for mathematical modeling. The same test will yield different results in the same person in different emotional states \cite{huth2016natural}. This variability extends right down to color perception and is even often used to determine emotional state \cite{maule2023development}. It is often impossible to overload the subject with too many tests.
It is also important that the tests used often overlap, as they are based on the same "biological substrate"—the human brain. The tasks of each psychological and psychophysiological test activate a large, but limited, number of brain regions in various combinations. Therefore, it is impossible to definitively state that each test reveals the activity of a specific brain region.
As an example, we can cite several facts. Arithmetic tests are often used to determine cognitive functions. Neuroimaging studies have demonstrated a clear involvement of the posterior parietal regions, particularly the intraparietal sulcus, angular gyrus, and supramarginal gyrus, which is part of the inferior parietal lobule, in the processing of numerical information and counting. In addition, the prefrontal, occipitotemporal, and hippocampal regions are involved in arithmetic calculations and counting \cite{butterworth2011dyscalculia,mccaskey2020persistent}. The hippocampus is also involved in orientation processes and in working and long-term memory \cite{duff2020semantic}. There is strong evidence of a relationship between the development of the cerebellum, whose activity is traditionally associated with balance, and the prefrontal cortex, as well as the significant role of the cerebellum in cognitive development \cite{zhang2023cerebellum}. Thus, assessing balance ability, as an indicator of brain health, activates brain areas that control attention and memory, help process information, and make decisions. There are many other similar examples. To overcome this limitation, it is necessary to isolate from each test as many different indicators as possible that can more accurately reflect narrower patterns.

Finally, any mathematical model requires a large sample size. Experimenters face limited opportunities to collect a representative sample. Moreover, data collection is often complicated by the fact that people may approach the proposed test set with varying degrees of thoroughness. This can lead to omissions, inaccuracies, selective incomplete completion, rapid, formal completion, and so on. These circumstances also limit the reliability of the obtained data. And this limitation cannot always be overcome by collecting a large sample. In some cases, the sample size is limited, requiring other approaches to address reliability issues.
Based on the presented facts, in this article, we present a comparative analysis of methods for constructing a mathematical model for determining cognitive age, taking into account all these limiting factors.

\section{Methods}
\label{sec:methods}

\subsection{Data collection}

During data collection, subjects were asked to complete a questionnaire and psychophysiological tests using the data collection system https://dev.cogni-test.ru/, which is accessible online. The test data collection system is a web application. It allows for individual users or groups of subjects to complete tests. The system does not allow for random ordering—users complete tests strictly in a predetermined sequence. The system does not collect or store any personal data. Participants complete tests anonymously, using only a numeric identifier (user\_id), which is not linked to real names, email addresses, or other personal information. Data (answers and reaction times) are stored in a SQLite database and analyzed separately from the server as part of scientific research.

All tasks (except instructions) are timed—from 3 to 15 seconds. Time is tracked using JavaScript on the client. The transition to the next test is automatic, with no option to return. The project logic sets a session limit of 15 minutes for completing all tests. If a user becomes "stuck" (for example, by leaving for a long time), access to the current tests is blocked after 15 minutes. Any attempt to submit a response or proceed further will result in redirection to the authorization page.

Thus, the system provides a rigid, linear structure: registration, authorization → completion → completion. There is no "exit" or progress review—the system is designed for a single, complete completion of the series within the allotted time. Data from the database can be downloaded in CSV format.

This study was conducted according to the guidelines of the Declaration of Helsinki and approved by the Ethics Committee of Lobachevsky University (Protocol number 3 from 8 April 2021).

\subsection{Description of tests}

The table \ref{tab:tests} provides a brief description of the tests and the variables for each subject.

\begin{table}[h!]
	\centering
	\begin{tabular}{|p{3.5cm}|p{4cm}|p{4cm}|}
		\hline
		\textbf{Test Name} & \textbf{Description} & \textbf{Variables} \\ \hline
		
		Reaction Speed Test &
		Participants judge 10 arithmetic expressions as correct or incorrect by pressing green or red buttons. &
		Average time, time per attempt, variance, overall accuracy, accuracy on correct/incorrect inequalities. \\ \hline
		
		Verbal Memory and Working Memory Capacity Test &
		Volunteers memorize 6 words, then identify words shown on screen as from the list or not. &
		Average time, time per attempt, variance, overall accuracy. \\ \hline
		
		Decision-Making Test (Stroop Test) &
		Participants select colors based on letter color or word meaning to test cognitive switching and conflict resolution. &
		Average time, time per decision, variance, overall accuracy, correctness of meaning and color responses, decision times by section. \\ \hline
		
		Spatial Perception Test &
		Respond to schematic swallow’s "flight" direction on colored backgrounds requiring same or opposite direction responses. &
		Average time, time per attempt, variance, overall accuracy, accuracy for "flying from" and "flying to" directions, decision times for each. \\ \hline
		
		Verbal Function Test (Munsterberg Test) &
		Search for nouns hidden in random letter sequences within 1 minute, recording number and proportion found. &
		Average time, time per word, variance, number of words found, proportion of words found. \\ \hline
		
		Color Campimetry Test &
		Identify when animal shapes become visible or invisible against changing hues by pressing buttons. &
		Average times and variances for stages 1 and 2, including times for figure visibility and invisibility. \\ \hline
		
	\end{tabular}
	\caption{Psychophysiological with descriptions and variables}
	\label{tab:tests}
\end{table}

\subsection{Machine learning algorithms}
To predict the subject's age based on the results of psychophysiological tests, the following regression algorithms were used: Random Forest Regressor, Extra Trees Regressor, Gradient Boosting Regressor, Support Vector Regression (SVR), Linear Regression, eXtreme Gradient Boosting (XGBoost), Light Gradient Boosting Machine (LightGBM), Lasso Cross-Validated (LassoCV), Ridge Cross-Validated (RidgeCV), Elastic Net Cross-Validated (ElasticNetCV), Adaptive Boosting Regressor (AdaBoost) and Bagging Regressor (Bagging). The best algorithm was selected based on the values of the determination coefficient and mean absolute error metrics.

\section{Results}
\label{sec:results}

\subsection{Data preprocessing}

The dataset contained 44 variables, one of which determined the unique subject number, one determined age, and one was categorical and determined gender. The remaining variables had numerical values. The data contained a slight gender imbalance (65 percent of the data were female and 35 percent male). 

\begin{figure}[h]
	\centering
	\includegraphics[width=0.65\textwidth]{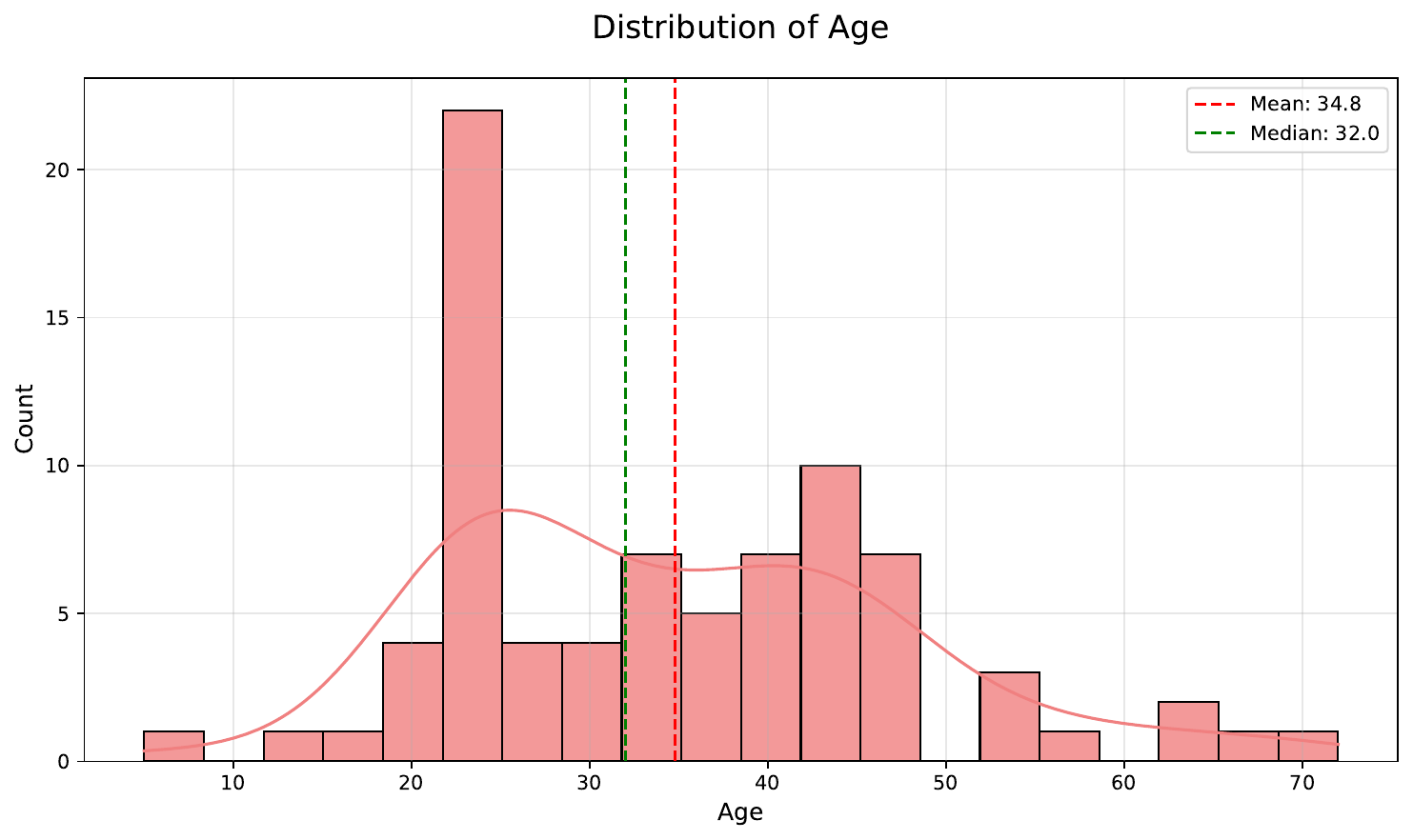}
	\caption{Distribution of the target variable "age".}
	\label{fig:rasp_age}
\end{figure}

The age distribution (Fig. \ref{fig:rasp_age}) shows that the mean age is 34, and the median is 32. Distributions of the remaining variables by test are presented in the Appendix (from Fig. \ref{fig:raspr_math} to Fig. \ref{fig:raspr_camp}).

The distribution of the variables shows that their values can differ by orders of magnitude, requiring a standardization procedure (Z-score normalization) to assess outliers. This data preprocessing method allows all numerical variables to be converted to a uniform scale using the formula:

\begin{equation}
	z = \frac{x - \mu}{\sigma}
\end{equation}

where:
\begin{itemize}
	\item $z$ is the standardized value
	\item $x$ is the original value of the variable
	\item $\mu$ is the sample mean of the variable
	\item $\sigma$ is the standard deviation of the variable
\end{itemize}

This procedure is available from the standard methods of the scikit-learn library.

\begin{figure}[h]
	\centering
	\includegraphics[width=0.95\textwidth]{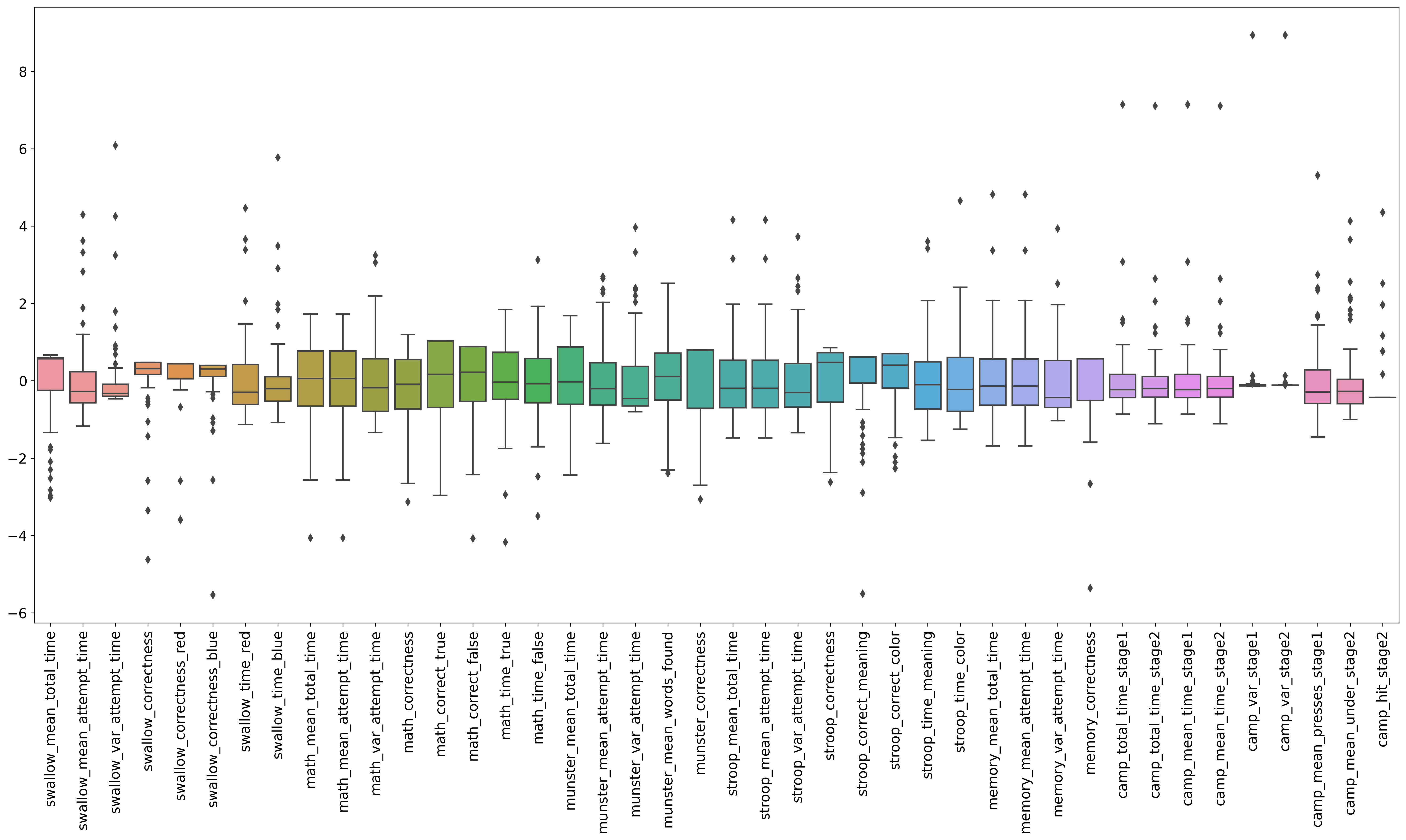}
	\caption{Boxplot plot of data variables.}
	\label{fig:boxplotvar}
\end{figure}

Data standardization and plotting a boxplot showing the distribution, median, quartiles, and outliers for the variables (Fig. \ref{fig:boxplotvar}) revealed that all variables in the data contain outliers. Moreover, outliers are present in 65 percent of the data, making it impossible to simply exclude them.   

To combat outliers, the winsorization procedure \cite{wilcox2011introduction,huber2011robust} was used. This is a statistical method for handling outliers in which extreme values are replaced by certain percentiles, which reduces the influence of outliers on the analysis results.

\subsubsection*{Mathematical description}

Let $X = \{x_1, x_2, \dots, x_n\}$ be the original sample of numerical values. We denote:

\begin{itemize}
	\item $q_\alpha$ — $\alpha$-quantile of the distribution
	\item $q_{1-\beta}$ — $(1-\beta)$-quantile of the distribution
	\item $\alpha$ — lower cutoff limit (usually 0.05)
	\item $\beta$ — upper cutoff limit (usually 0.05)
\end{itemize}

Then the winsorized sample $X_w$ is defined as:

\[
x_w^{(i)} = 
\begin{cases}
	q_\alpha, & \text{if } x_i \leq q_\alpha \\
	x_i, & \text{if } q_\alpha < x_i < q_{1-\beta} \\
	q_{1-\beta}, & \text{if } x_i \geq q_{1-\beta}
\end{cases}
\]

\subsection{Removing multicollinearity from data}

\begin{figure}[h]
	\centering
	\includegraphics[width=0.95\textwidth]{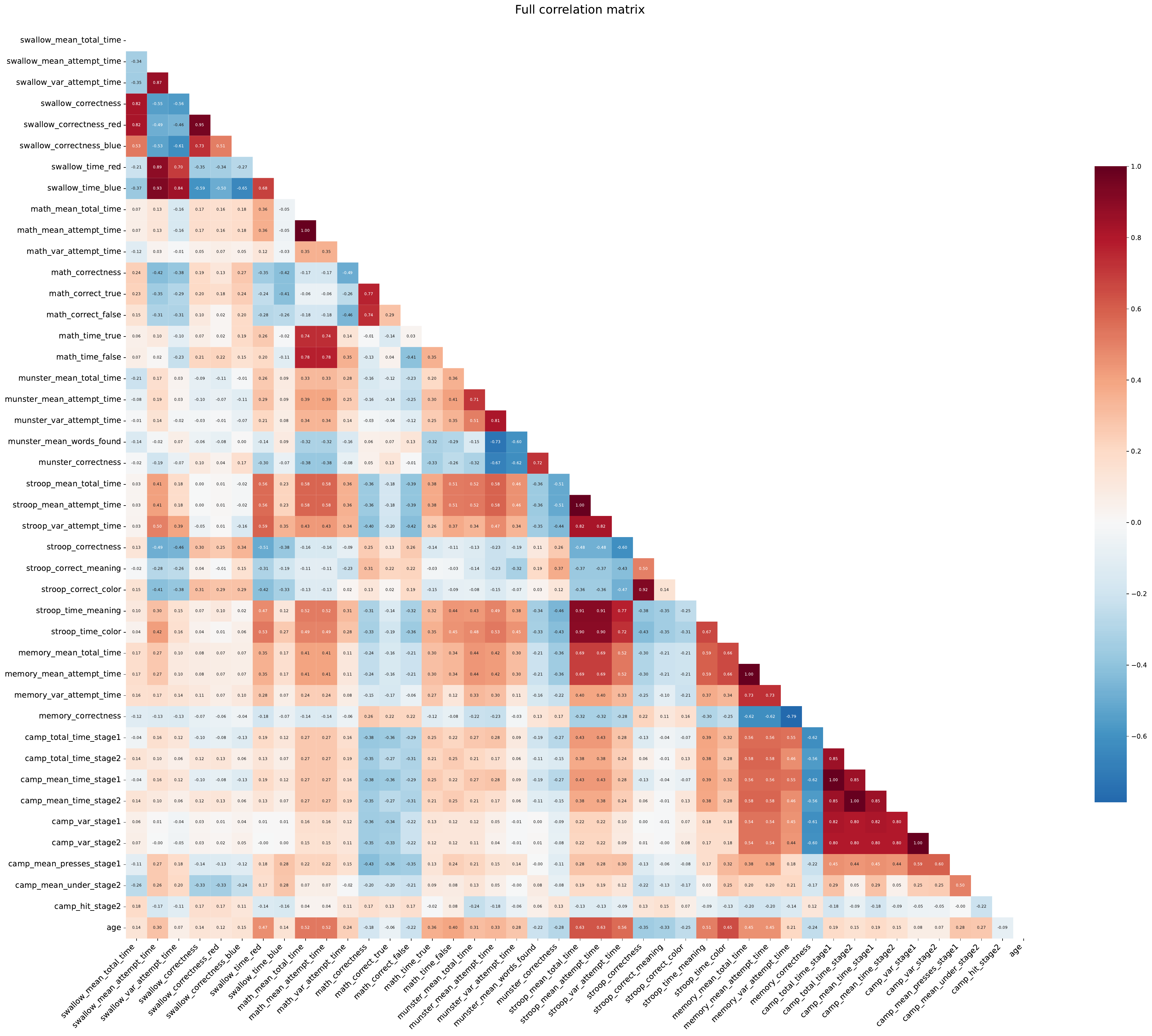}
	\caption{Triangular correlation matrix of data variables.}
	\label{fig:corr_matrix}
\end{figure}

To check for multicollinearity, we need to transform categorical features (in our case, "gender") into numerical values. For this, we'll use the $LabelEncoder$ class from the $scikit-learn$ library. Since we have 44 variables, including the target, we'll use a triangle to better visualize the collinearity of the variables (Fig. \ref{fig:corr_matrix}).

The Fig. \ref{fig:corr_matrix} presents a triangular correlation matrix heatmap summarizing the relationships among numerous cognitive and sensorimotor variables derived from various behavioral tasks, including Reaction Speed Test, Verbal Function Test, Verbal Memory and Working Memory Capacity Test, Decision-Making Test, Color Campimetry Test, and Spatial Perception Test, as well as the variable Age. Only the lower half of the matrix is visualized to avoid redundancy, since a correlation matrix is symmetric. The heatmap employs a diverging color scale ranging from negative to positive values, typically with blue for negative correlations and red or orange for positive correlations. The dataset includes around 40 features grouped by task type. Within each cognitive test, the measured variables show strong positive correlations, often in the range of 0.8 to 0.95, indicating high internal consistency among metrics within the same task. 

For example, the mean, variance, and attempt times in the Reaction Speed Tests and Decision-Making Tests correlate strongly with each other, suggesting that these features capture the same underlying cognitive speed component. This coherence is also observed in the Spatial Perception Test metrics, where time-based features cluster together closely. Across different cognitive tests, the correlations are moderate (approximately 0.5–0.7), reflecting the presence of a shared latent factor related to general cognitive speed or executive function. For instance, reaction time variables from Reaction Speed Tests, Decision-Making Test, and Verbal Memory and Working Memory Capacity Test tasks exhibit moderate positive correlations, indicating that individuals who are slower in one task tend to be slower across other tasks as well. However, accuracy and correctness measures show weaker or even negative correlations across tasks, implying that they capture task-specific processes rather than a unified construct. A key observation involves the relationship between task performance metrics and the Age variable. Age correlates positively with various timing measures and negatively with correctness scores. Typically, timing features such as total or mean attempt time correlate around 0.4 to 0.7 with age, while correctness metrics exhibit correlations between –0.3 and –0.6. This pattern indicates that older participants tend to perform tasks more slowly and slightly less accurately, which is consistent with known effects of aging on psychomotor and cognitive processing speed. Another noteworthy aspect is the presence of very strong correlations between some feature pairs (greater than 0.9), which signals significant redundancy and potential multicollinearity. In the context of regression or predictive modeling, such highly interrelated variables would need to be handled carefully, for example by applying feature selection, variance inflation factor (VIF) analysis. Overall, this correlation matrix highlights several important patterns: cognitive task measures show high internal coherence within each domain, moderate shared variance across tasks pointing to global cognitive processing effects, and clear age-related trends in both speed and accuracy. The visualization thus provides a comprehensive overview of how diverse cognitive and motor features relate to each other and how they jointly reflect broader aspects of cognitive aging.

We plot the most highly correlated pairs of features (correlation values greater than 0.5) in a matrix plot showing the pairwise relationships between variables in the data set (Fig. \ref{fig:pairplot}). Of all the features, eight were found to be the most correlated: 

\begin{enumerate}
	\item $swallow\_correctness$
	\item $camp\_mean\_time\_stage2$
	\item $math\_mean\_attempt\_time$
	\item $stroop\_mean\_attempt\_time$
	\item $stroop\_correctness$
	\item $stroop\_time\_meaning$
	\item $swallow\_mean\_attempt\_time$
	\item $memory\_mean\_attempt\_time$
\end{enumerate}

These variables reflect the results of several tests: Verbal Memory and Working Memory Capacity Tests, Decision-Making Tests, Color Campimetry Tests, and Spatial Perception Tests. The results of the reaction speed test were the least correlated relative to the other features. The results of the reaction speed test were the least correlated relative to the other features.The plot reveals strong positive correlations among time-based features and negative correlations between timing and correctness measures.

To reduce multicollinearity in the data, we use the Variance Inflation Factor (VIF) \cite{murray2012variance,liao2012variance}. This is a statistical metric that measures the degree of multicollinearity in regression analysis. VIF shows how much the variance of the regression coefficient is increased due to a linear relationship between the predictors.

The Variance Inflation Factor for the kth predictor is defined as:

$$
\text{VIF}_k = \frac{1}{1 - R^2_k}
$$

where $R^2_k$ — coefficient of determination of the regression model:
$$
X_k = \beta_0 + \beta_1 X_1 + \cdots + \beta_{k-1} X_{k-1} + \beta_{k+1} X_{k+1} + \cdots + \beta_p X_p + \varepsilon
$$

The Variance Inflation Factor (VIF) results indicate the degree of multicollinearity among your features, which affects the reliability of regression coefficients. Features with VIF above 10 are considered highly collinear, leading to unstable and inflated coefficient estimates. The removed features have extremely high VIF values, ranging from 11 to as high as 390, signaling severe multicollinearity issues. These include several Stroop and Math timing variables, memory times, and correctness scores, which likely share overlapping information.

In contrast, the remaining features—swallow\_time\_red (VIF = 6.61), munster\_mean\_attempt\_time (VIF = 5.44), and stroop\_var\_attempt\_time (VIF = 5.21)—have moderate multicollinearity. Their VIF values are below the critical threshold of 10, making them more stable and reliable predictors. Retaining these helps reduce redundancy while preserving diverse aspects of cognitive performance in the model.

\begin{figure}[h]
	\centering
	\includegraphics[width=0.95\textwidth]{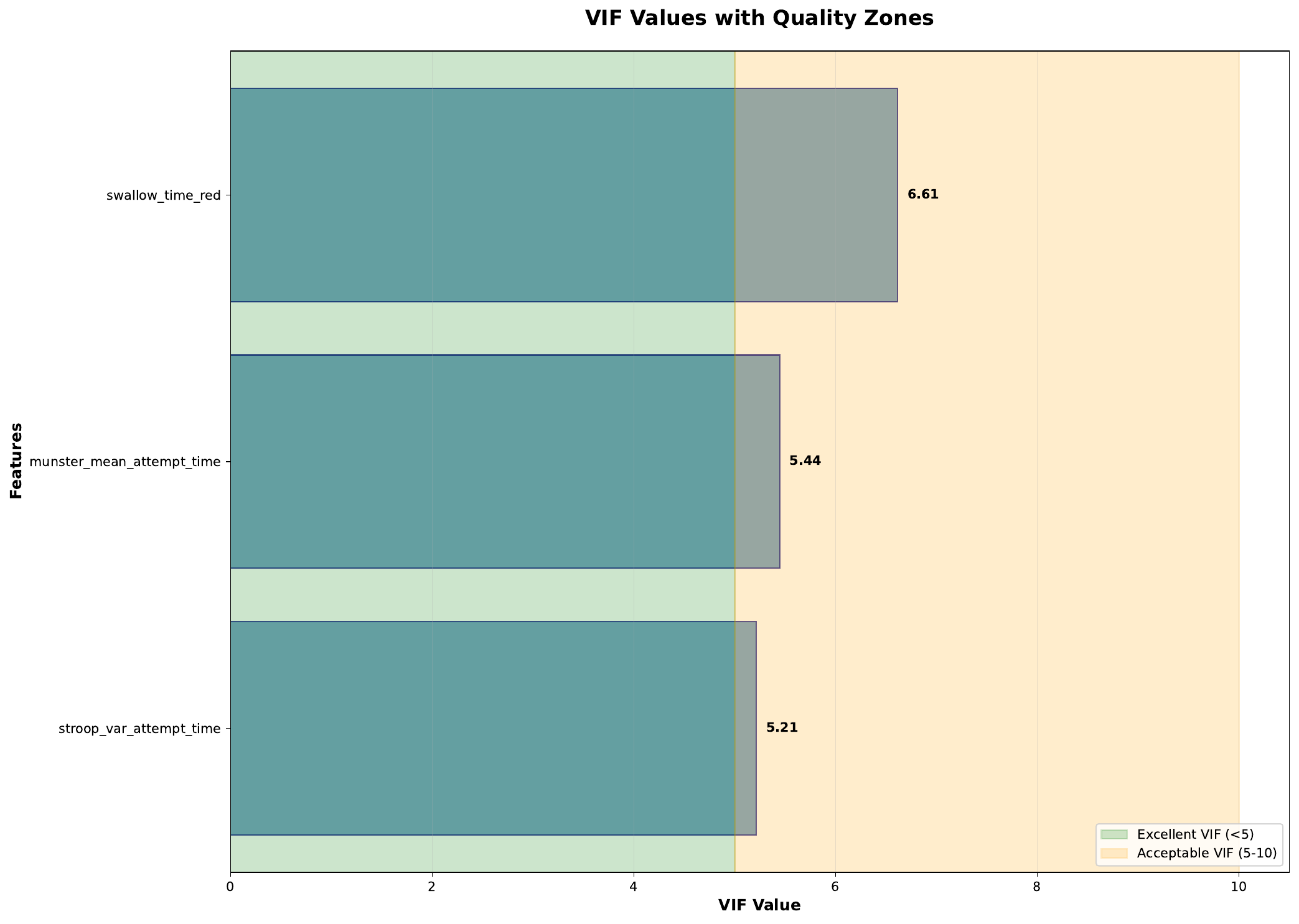}
	\caption{Selected 12 variables with the best VIF value in the range from 0 to 10.}
	\label{fig:VIF_top}
\end{figure}

Based on the analysis results, 3 features with the lowest VIF in the range from 0 to 10 were selected (Fig. \ref{fig:VIF_top}). The area with the best VIF value (less than 5) is marked in green, while those with a satisfactory value suitable for use in the model (from 5 to 10) are marked in yellow (the remaining 3 features).

\subsection{Training and testing models}

To select the optimal model, several regressor models were used:  Random Forest Regressor, Extra Trees Regressor, Gradient Boosting Regressor, Support Vector Regression (SVR), Linear Regression, eXtreme Gradient Boosting (XGBoost), Light Gradient Boosting Machine (LightGBM), Lasso Cross-Validated (LassoCV), Ridge Cross-Validated (RidgeCV), Elastic Net Cross-Validated (ElasticNetCV), Adaptive Boosting Regressor (AdaBoost) and Bagging Regressor (Bagging). The test sample was 20 percent, and the training sample was 80 percent due to the small dataset.

\begin{figure}[h]
	\centering
	\includegraphics[width=0.95\textwidth]{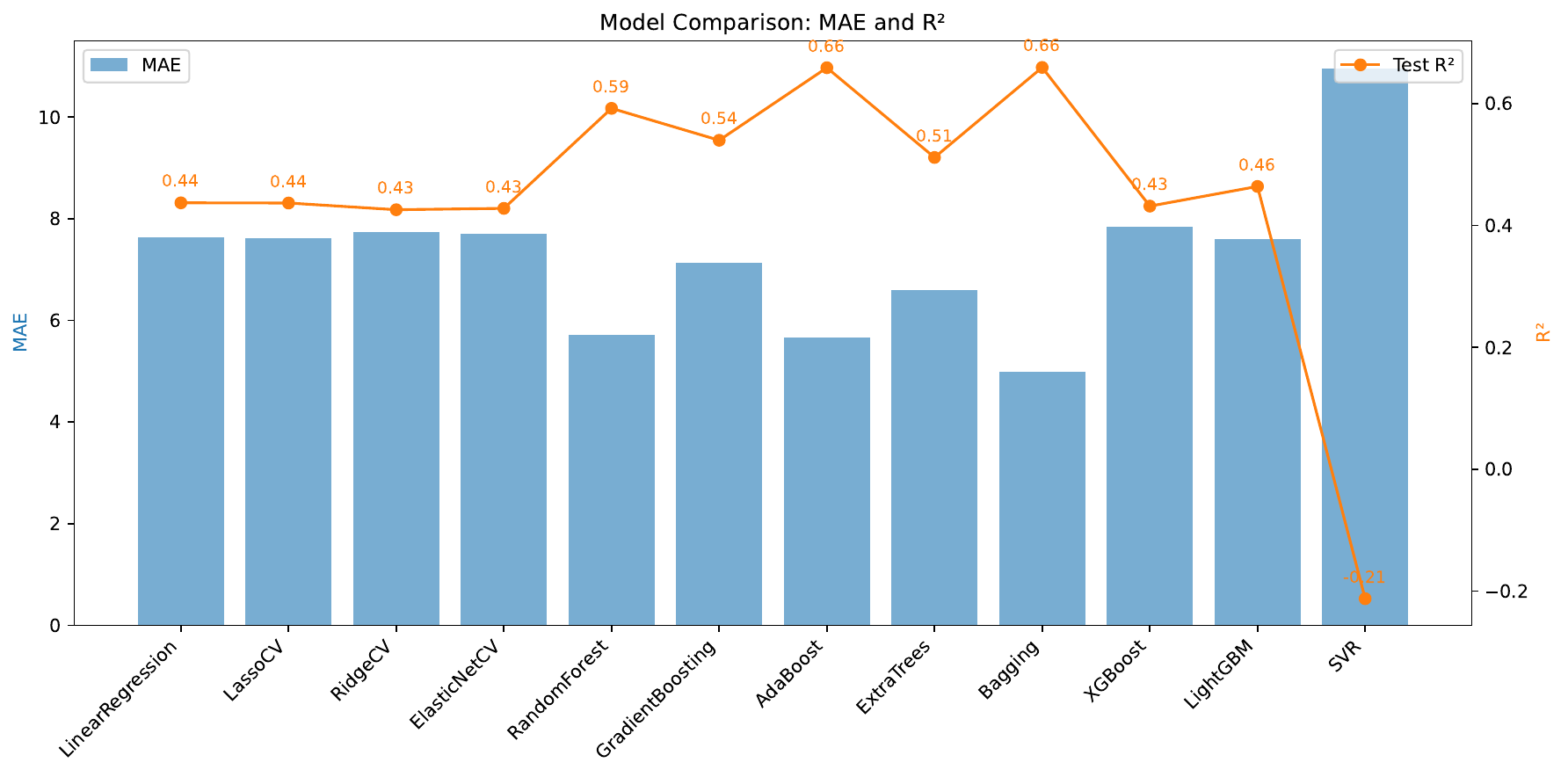}
	\caption{Comparison of models on test data by MAE and $R^2$.}
	\label{fig:model_select}
\end{figure}

Based on a comparative analysis of the machine learning results for the regression problem, the following conclusions can be drawn about the performance of the various algorithms (Fig. \ref{fig:model_select}). Model evaluation revealed clear performance differences between linear and ensemble approaches. Linear models, including Linear Regression, LassoCV, RidgeCV, and ElasticNetCV, all produced similar results with mean absolute errors (MAE) around 7.6 and test $R^2$ values near 0.44, suggesting limited predictive power and poor generalization (cross-validation $R^2$ values close to zero or negative). Ensemble models, on the other hand, performed substantially better. AdaBoost and Bagging achieved the highest test $R^2$ values (0.66) and the lowest MAEs (5.66 and 4.99 respectively), indicating strong predictive accuracy. RandomForest and GradientBoosting also performed reasonably well ($R^2$ between 0.54 and 0.59), while XGBoost and LightGBM showed only moderate results ($R^2$ around 0.45). SVR performed poorly, with a negative $R^2$, confirming that it was not well-suited to this dataset.

After controlling for multicollinearity, only three stable and informative predictors remained. Linear models were unable to capture the complexity of the data, whereas nonlinear ensemble methods, particularly AdaBoost and Bagging, provided the most accurate and robust performance. This suggests that the underlying relationships between the remaining features and the target variable are nonlinear and best modeled with ensemble-based approaches.

\begin{figure}[h]
	\centering
	\textit{(a)}\includegraphics[width=0.95\textwidth]{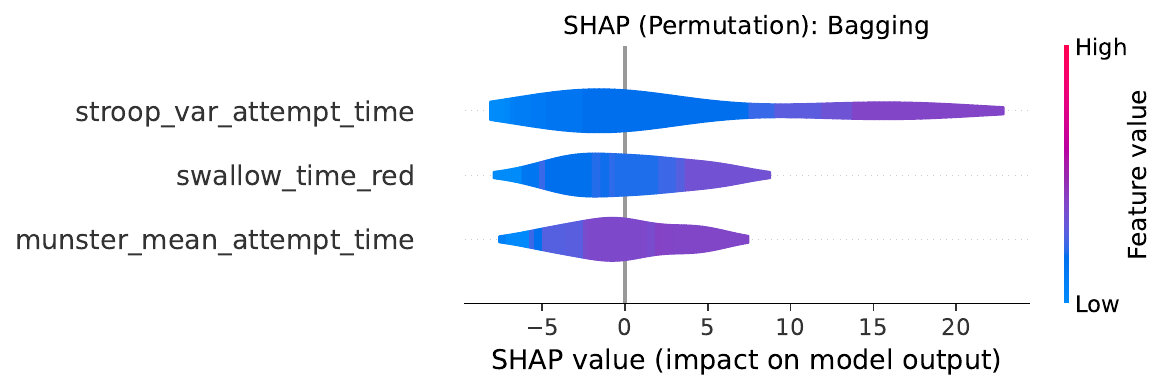} \\
	\quad \textit{(b)} \includegraphics[width=0.95\textwidth]{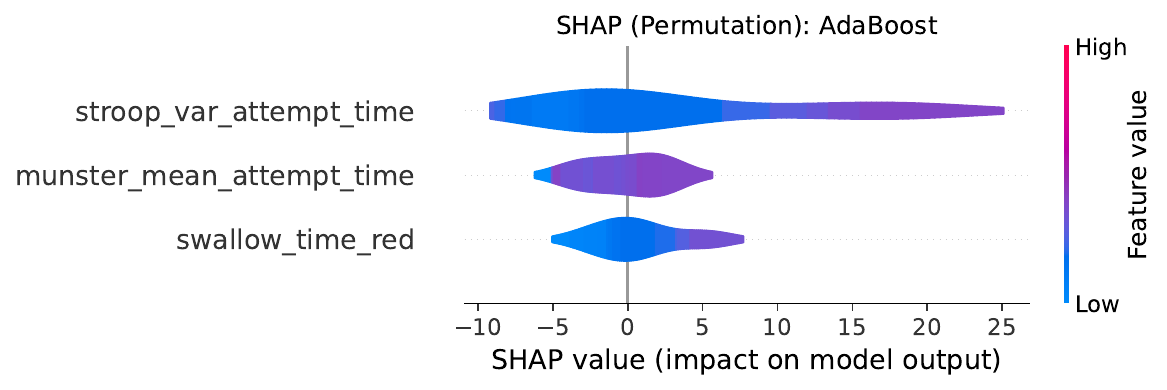}
	\caption{SHAP analysis for a) Bagging model and b) AdaBoost model.}
	\label{fig:model_shap}
\end{figure}

The SHAP (permutation) plots for the Bagging and AdaBoost models illustrate how each of the three retained features—$swimlow\_time\_red$, $munster\_mean\_attempt\_time$, and $stroop\_var\_attempt\_time$—affects the model's predictions (Fig. \ref{fig:model_shap}). In these plots, the x-axis represents the SHAP value, indicating the magnitude and direction of each feature's contribution to the prediction, and the color gradient (low to high) displays the actual feature value. The y-axis lists the features ordered by their overall importance to the model. SHAP analysis explains how each feature influences a model’s predictions by assigning an importance value (SHAP value) to every feature for each prediction. Positive SHAP values indicate a feature increases the predicted outcome, while negative values decrease it.

In the Bagging model, $stroop\_var\_attempt\_time$ shows the largest range of SHAP values, indicating that this feature is the most influential. Changes in this variable have the greatest impact on the model's output. $munster\_mean\_attempt\_time$ has a moderate impact on SHAP, making both positive and negative contributions to predictions, indicating that the model accounts for nonlinear relationships between this feature and the outcome. $swallow\_time\_red$ has the smallest range of SHAP, indicating that it plays a weaker, albeit significant, role in prediction.

In the AdaBoost model, the pattern of feature importance remains similar. $stroop\_var\_attempt\_time$ is also the most influential predictor, demonstrating the largest range of SHAP values (approximately -10 to +25), indicating a strong and asymmetric influence of this feature on the model's results. The remaining features have a similar pattern of influence as in the Bagging model. The broader and more uneven distribution of SHAP in the AdaBoost model suggests that it accounts for stronger nonlinear and interactive effects than bagging.
These SHAP plots confirm that the remaining features significantly influence the predictions and that both ensemble models capture complex nonlinear relationships within the data.





\section{Discussion}
\label{sec:discussion}

In our study, we analyzed data obtained from psychophysiological tests—reaction time, spatial perception, color campimetry, and the Stroop and Mansberg tests—to construct regression models. A characteristic feature of such data is its high sensitivity to psychophysiological, emotional, and cognitive personality traits. Reaction time characteristics, the influence of fatigue, anxiety, and other factors lead to a large number of outliers, requiring robust data processing methods.

In the original data, all variables contained outliers across 43 variables, and significant multicollinearity was observed. Only 3 variables had acceptable correlation levels (VIF from 0 to 10). Winsorization was used to handle outliers, limiting extreme values and reducing distortions caused by anomalies. This approach stabilized the feature distribution, reduced variance, and increased the reliability of model estimates even with small sample sizes and high individual variability.

Several regression algorithms were compared to identify the best-performing model. Linear models (Linear Regression, LassoCV, RidgeCV, ElasticNetCV) showed limited predictive power ($MAE = 7.6$, $R^2 = 0.44$), while ensemble methods performed much better. AdaBoost and Bagging achieved the highest accuracy ($R^2 = 0.66$, $MAE = 5.66 and 4.99$), followed by Random Forest and Gradient Boosting ($R^2 = 0.54-0.59$). After removing multicollinearity, three key features remained: $swallow\_time\_red$,$ munster\_mean\_attempt\_time$, and $stroop\_var\_attempt\_time$, indicating nonlinear relationships best captured by ensemble models.

SHAP analysis showed that $stroop\_var\_attempt\_time$ had the strongest influence on predictions, while the other two features had moderate effects. AdaBoost displayed a wider SHAP range, suggesting stronger nonlinear and interaction effects. Overall, ensemble models—especially AdaBoost and Bagging—provided the most accurate and interpretable results.

Thus, an integrated approach, including winsorization for outlier handling, multicollinearity control using VIF, and the use of ensemble algorithms, improves the reproducibility and interpretability of models when analyzing complex psychophysiological data with small sample sizes and significant individual variability.

\section{Conclusion}
\label{sec:conclusion}

The conducted research has shown that when working with small volumes of psychophysiological data containing high multicollinearity and outliers, the use of complex preprocessing, including winsorization and feature selection based on VIF, plays a key role in improving the quality of regression models. The specific characteristics of data obtained from subjects in psychophysiological tests of reaction time, spatial perception, and cognitive functions (such as the Stroop and Mansberg tests) require special attention to the processing of outliers and feature stability due to high variability and the influence of psychophysiological factors.

The results showed that ensemble models, in particular AdaBoost and Bagging, provide more accurate and robust predictions compared to classical linear models, which is consistent with modern concepts regarding the need to use robust algorithms for psychophysiological data analysis. Thus, the combination of winsorization, multicollinearity control, and ensemble algorithms is an effective approach for analyzing and modeling complex psychophysiological data.

\section{Acknowledgements}

This research was funded by the Ministry of Science and Higher Education of the Russian Federation (project FSWR-2025-0009).

\appendix
\section[\appendixname~\thesection]{Additional graphs}

\begin{figure}[H]
	\centering
	\includegraphics[scale=0.45]{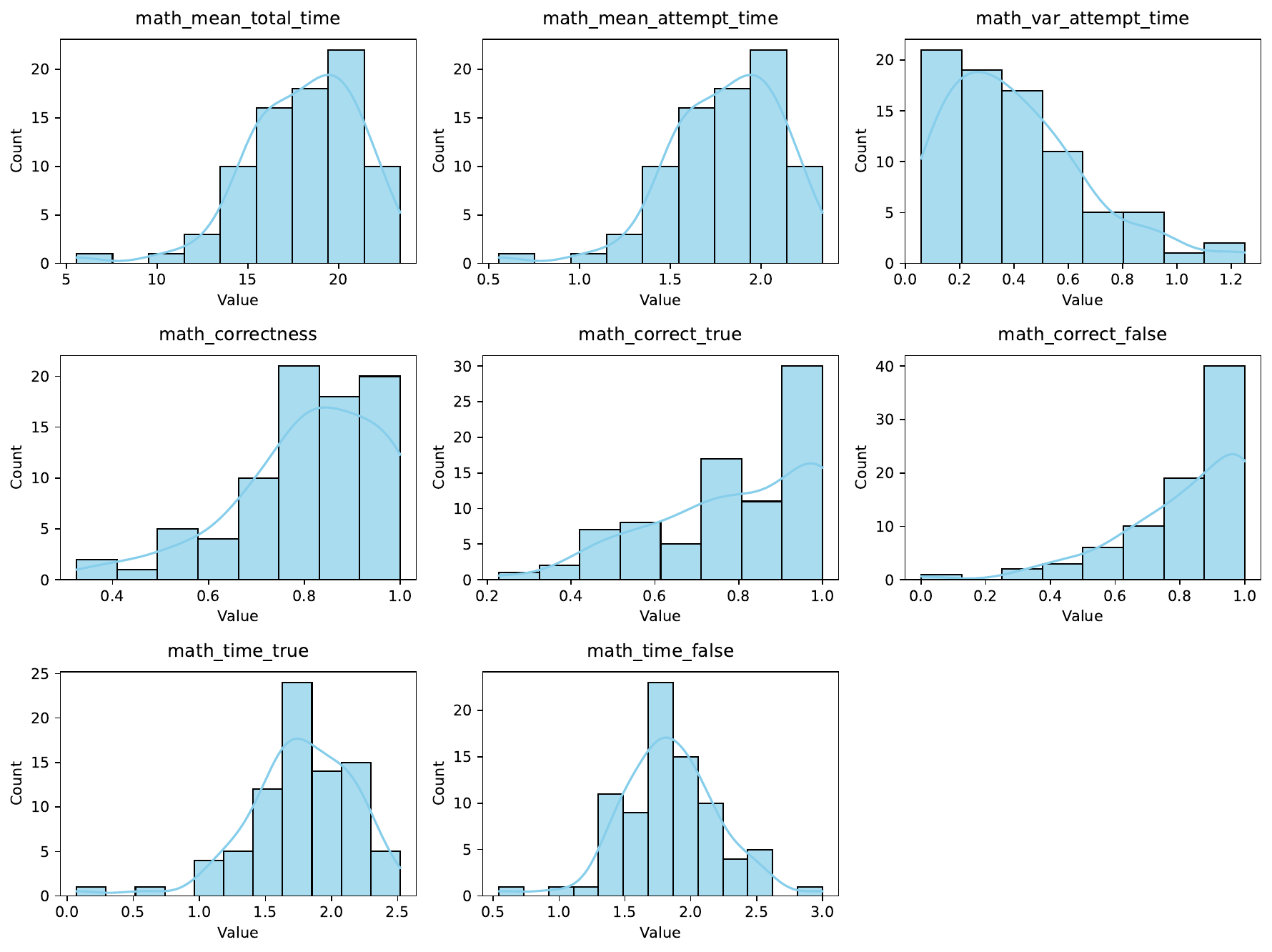}
	\caption{Distribution of variable values for the reaction speed test}
	\label{fig:raspr_math}
\end{figure}

\begin{figure}[H]
	\centering
	\includegraphics[scale=0.45]{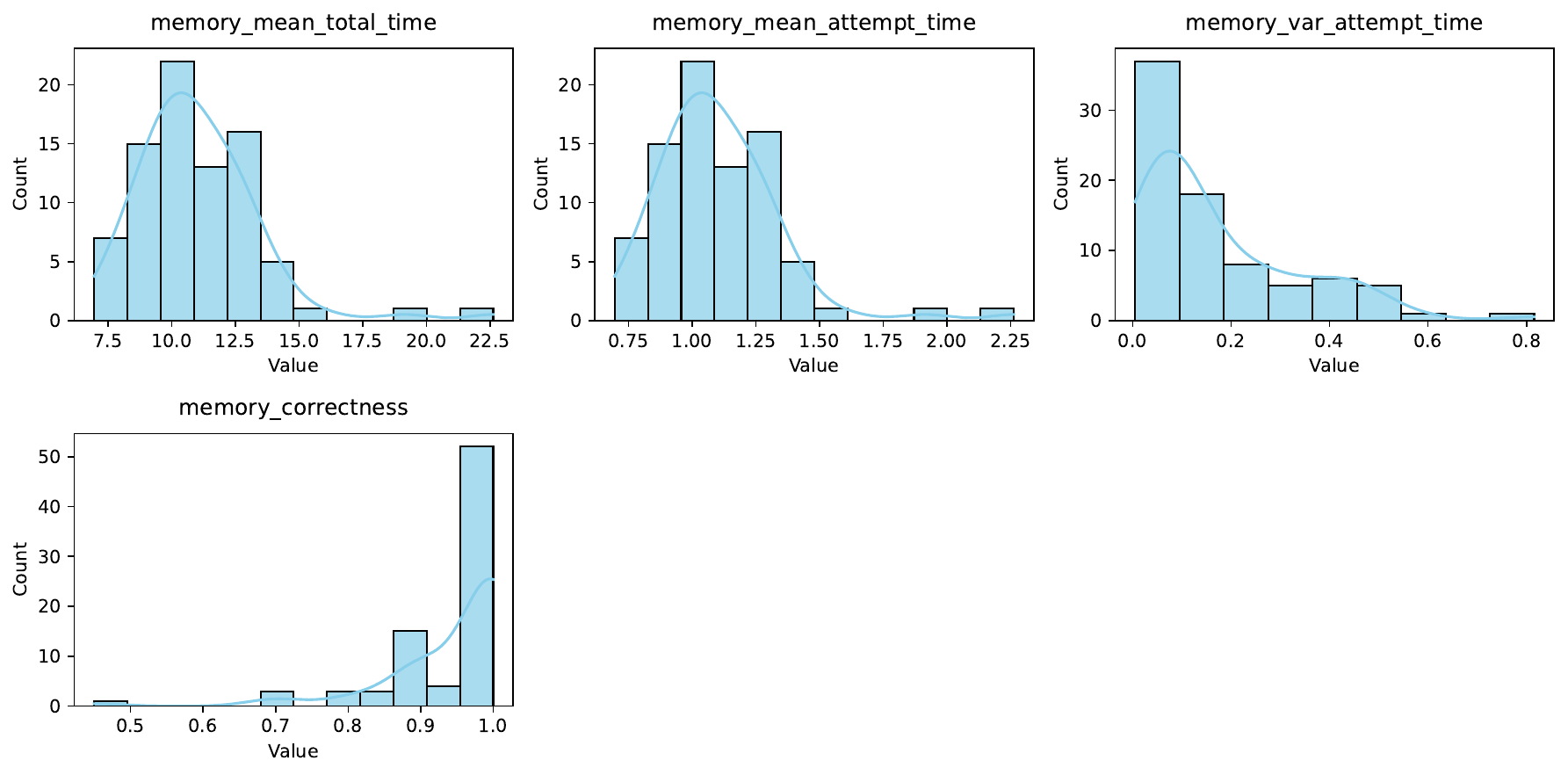}
	\caption{Distribution of variable values for the verbal memory test and working memory capacity}
	\label{fig:raspr_memory}
\end{figure}

\begin{figure}[H]
	\centering
	\includegraphics[scale=0.45]{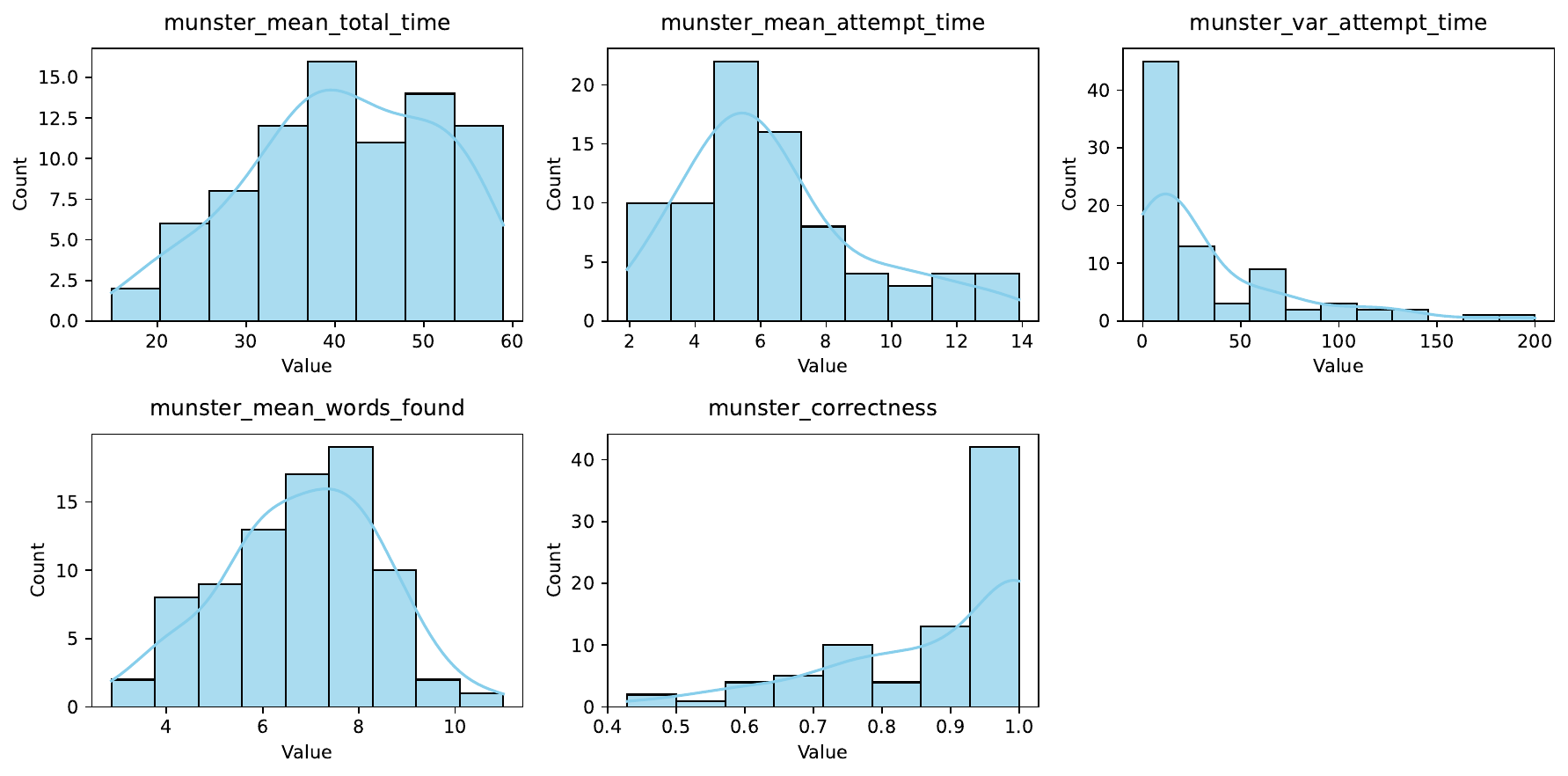}
	\caption{Distribution of variable values for the verbal function test}
	\label{fig:raspr_munster}
\end{figure}

\begin{figure}[H]
	\centering
	\includegraphics[scale=0.45]{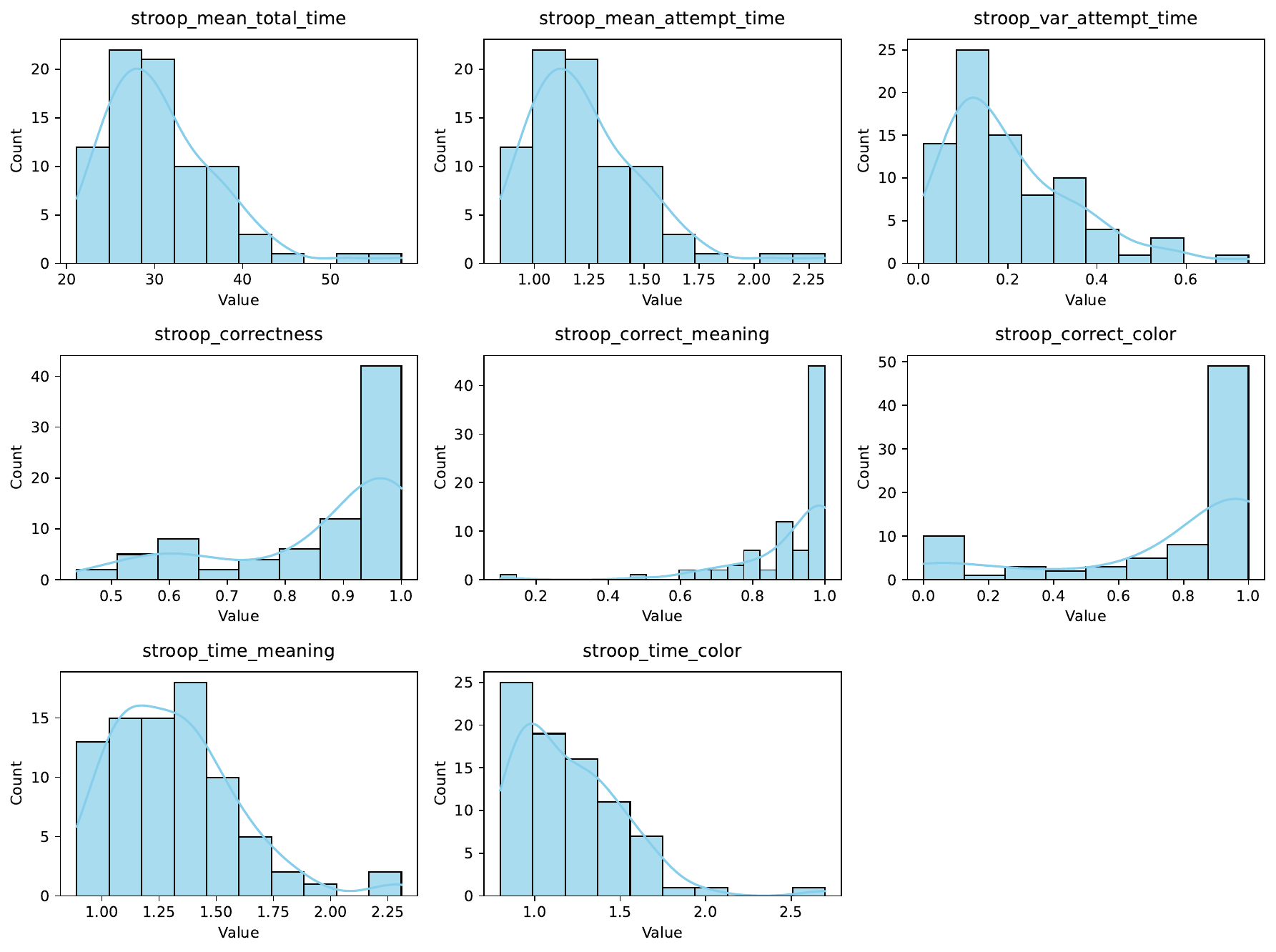}
	\caption{Distribution of variable values for the decision-making ability test}
	\label{fig:raspr_stroop}
\end{figure}

\begin{figure}[H]
	\centering
	\includegraphics[scale=0.45]{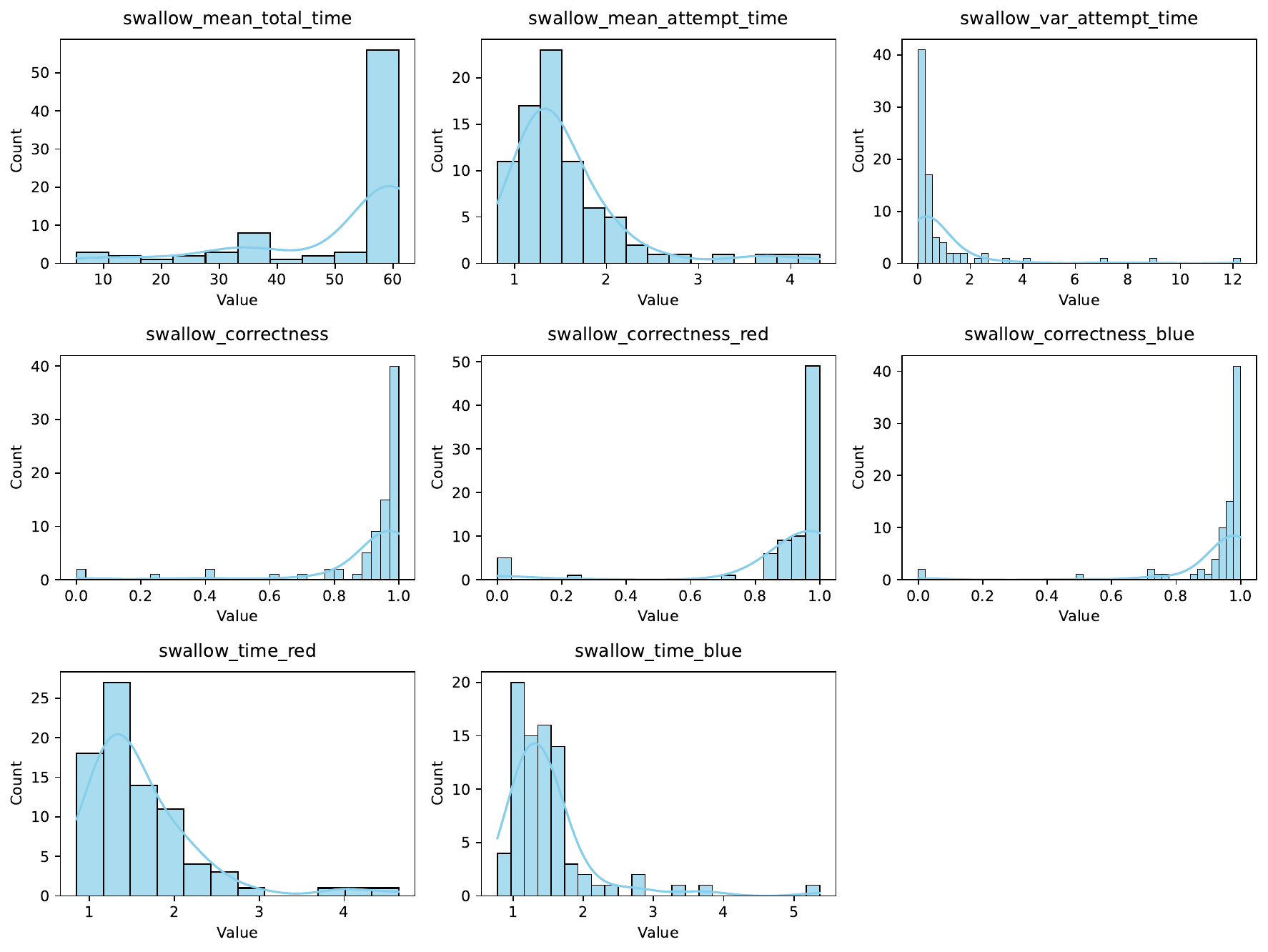}
	\caption{Distribution of variable values for the spatial perception test}
	\label{fig:raspr_swallow}
\end{figure}

\begin{figure}[H]
	\centering
	\includegraphics[scale=0.45]{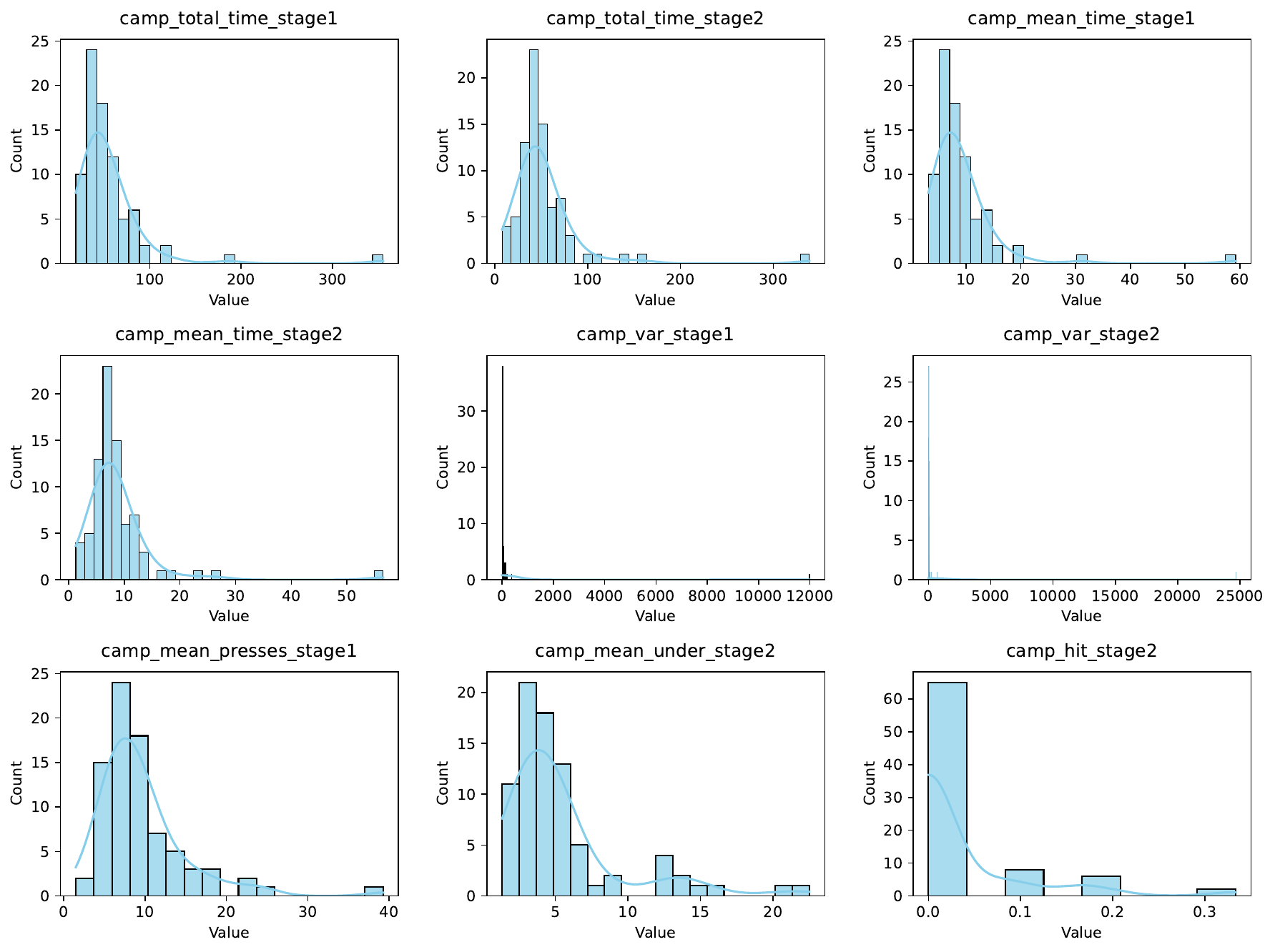}
	\caption{Distribution of variable values for the spatial perception test}
	\label{fig:raspr_camp}
\end{figure}


\begin{figure}[h]
	\centering
	\includegraphics[width=0.95\textwidth]{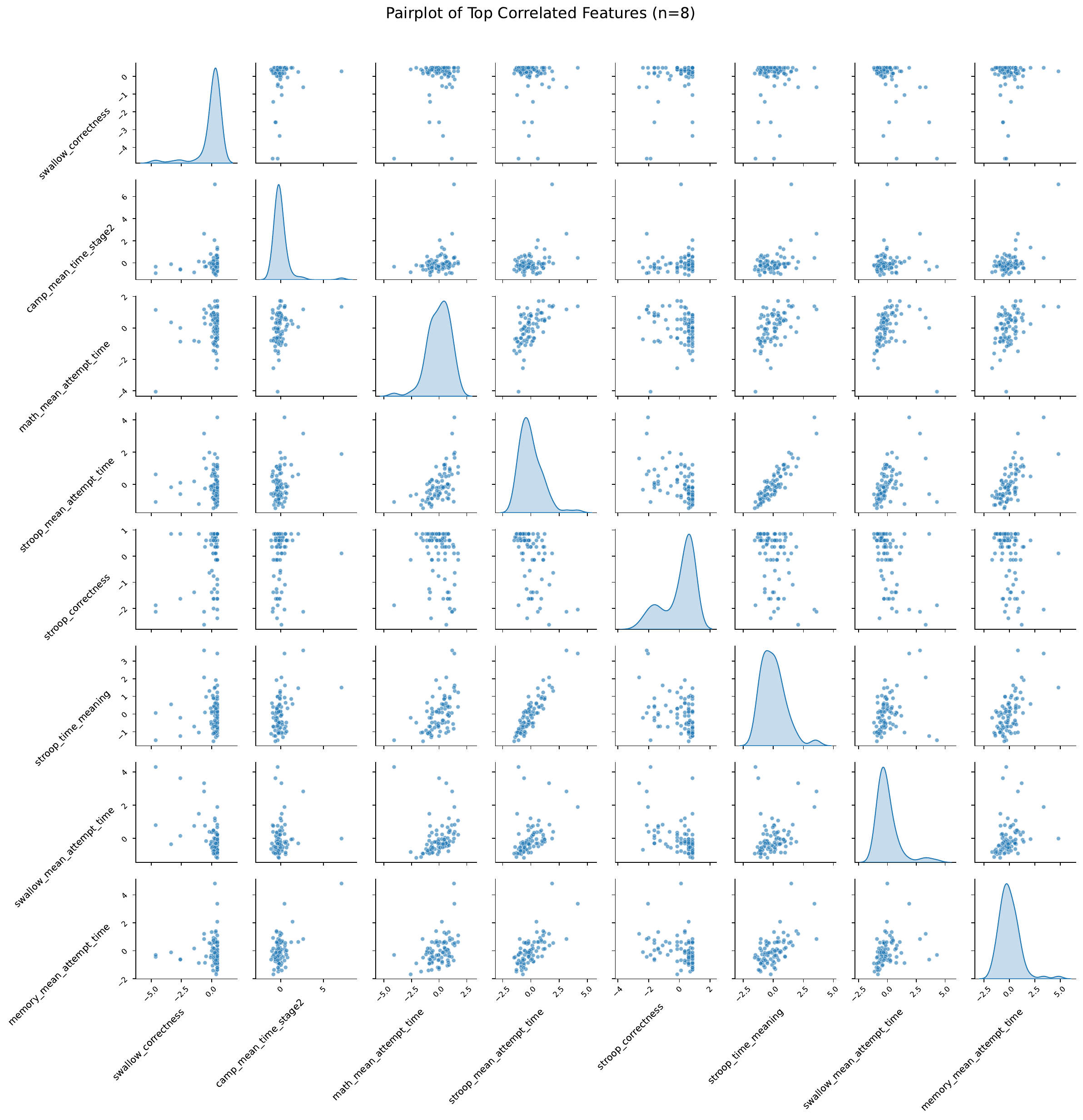}
	\caption{A matrix of plots showing the pairwise relationships between highly correlated variables in a data set.}
	\label{fig:pairplot}
\end{figure}

\end{document}